\documentclass[final,3p,times,10pt]{elsarticle}
\usepackage{graphicx,color,amsmath,amssymb}
\usepackage{bm}
\usepackage[colorlinks=true, pdfstartview=FitV, linkcolor=red, citecolor=blue, urlcolor=blue]{hyperref}

\advance\voffset -0.2in
\biboptions{sort&compress}

\begin{document}

\begin{frontmatter}
\title{A remark on the sign change of the four-particle azimuthal cumulant\\in small systems}

\author[agh]{Adam Bzdak}
\ead{bzdak@fis.agh.edu.pl}

\author[glm]{Guo-Liang Ma}
\ead{glma@sinap.ac.cn}

\address[agh]{AGH University of Science and Technology,
Faculty of Physics and Applied Computer Science,
30-059 Krak\'ow, Poland}

\address[glm]{Shanghai Institute of Applied Physics, 
Chinese Academy of Sciences, Shanghai 201800, China}

\begin{abstract}  
The azimuthal cumulants, $c_2\{2\}$ and $c_2\{4\}$, originating from the global conservation of transverse momentum in the presence of hydro-like elliptic flow are calculated. We observe the sign change of $c_2\{4\}$ for small number of produced particles. This is in a qualitative agreement with the recent ATLAS measurement of multi-particle azimuthal correlations with the subevent cumulant method.
\end{abstract}

\end{frontmatter}

\section{Introduction}

Experimental results from  
heavy-ion colliders indicate that a nearly perfect fluid is produced in high energy nucleus-nucleus (A+A) collisions~\cite{Adams:2005dq,Adcox:2004mh,Aamodt:2008zz}. One important evidence is the success of hydrodynamics in describing the collective flow phenomena observed in A+A, see, e.g.,~\cite{Florkowski:book,Teaney:2001av,Song:2007ux,Luzum:2008cw,Bozek:2009dw,Schenke:2010rr}.
The hydrodynamical models capture the main features of collective flow measured using different methods~\cite{Poskanzer:1998yz,Borghini:2000sa,Bilandzic:2010jr,Bhalerao:2003xf,Borghini:2004ke}. For example, the k-particle azimuthal cumulants, $c_n\{k\}$, are expected to measure the {\it real} collective flow $v_n$ by reducing non-flow effects~\cite{Borghini:2000sa,Bilandzic:2010jr}. The experimental results from the Large Hadron Collider (LHC) show that the elliptic flow coefficients obtained with four, six and eight-particle standard cumulant method are overlapping in both Pb+Pb and p+Pb collisions, indicating that the observed long-range (in rapidity) azimuthal correlations may be due to the same physical origin in both large and small systems~\cite{Khachatryan:2015waa,Aaboud:2017acw,Gajdosova:2017fsc}.

A new subevent cumulant method was recently developed to further suppress the non-flow contribution from jets~\cite{Jia:2017hbm}. The ATLAS measurement~\cite{Aaboud:2017blb} demonstrated that the two-subevent and three-subevent cumulants are less sensitive to short-range non-flow effects than the standard cumulant method. The three-subevent method shows that $c_2\{4\}$ in proton-proton and p+Pb collisions changes sign at lower multiplicity than the standard method, indicating that the long-range multi-particle azimuthal correlations persist to even lower multiplicities. On the other hand, many theoretical efforts have been made to understand these measurements, which are basically classified as final state~\cite{Bozek:2011if,Bzdak:2013zma,Shuryak:2013ke,Qin:2013bha,
Bozek:2013uha,Ma:2014pva,Bzdak:2014dia,Koop:2015wea,Shen:2016zpp,Weller:2017tsr,Song:2017wtw} 
or initial state phenomena~\cite{Dumitru:2010iy,Kovchegov:2012nd,Dusling:2013qoz,Skokov:2014tka,Dumitru:2014yza,Schenke:2015aqa,
Schlichting:2016sqo,McLerran:2016snu,Kovner:2016jfp,Dusling:2017aot}, see~\cite{Dusling:2015gta} for a recent review. 

In this paper we calculate the two-particle and the four-particle azimuthal cumulants 
\begin{equation}
c_{2}\{2\}=\left\langle e^{i2(\phi _{1}-\phi _{2})}\right\rangle ,
\label{eq:c22-def}
\end{equation}
\begin{equation}
c_{2}\{4\}=\left\langle e^{i2(\phi _{1}+\phi _{2}-\phi _{3}-\phi
_{4})}\right\rangle -2\left\langle e^{i2(\phi _{1}-\phi _{2})}\right\rangle
^{2},  \label{eq:c24-def}
\end{equation}
originating from the conservation of transverse momentum in the presence of
hydro-like elliptic flow.

Recently we calculated the effect of transverse momentum conservation (TMC) only \cite{Bzdak:2017zok}, 
and we observed that
\begin{equation}
c_{2}\{k\}\sim \frac{1}{N^{k}},
\end{equation}
with $c_{2}\{k\}>0$ for the calculated $k=2,4,6,8$.\footnote{For comparison, clusters decaying into $k$ particles 
result in $c_{2}\{k\}\sim 1/N^{k-1}$, see, e.g., Ref. \cite{Borghini:2000sa}.} Here $N$ is the number
of produced particles subjected to TMC. As shown in \cite{Bzdak:2017zok},
the contribution from TMC to  $(c_{2}\{k\})^{1/k}$ is of the order of a few percent even for a
relatively large number of particles. In this paper we extend our analysis 
and calculate analytically $c_{2}\{2\}$ and $c_{2}\{4\}$
originating from TMC applied to particles characterized by the hydro-like elliptic flow.
We observe that $c_{2}\{4\}$ changes sign for small $N$ in a qualitative
agreement with the recent ATLAS measurement of multi-particle azimuthal correlations 
with the subevent cumulant method \cite{Jia:2017hbm,Aaboud:2017blb}.

\section{Calculation}

We calculate the effect of TMC applied to particles characterized by the hydro-like elliptic flow. This can be modeled by a single particle distribution given by\footnote{We neglect $v_3$ which also contributes to $c_2\{2\}$ and $c_2\{4\}$ however, its effect is smaller than $v_2$.}
\begin{equation}
f(p,\phi )=\frac{g(p)}{2\pi }\left[ 1+2v_{2}(p)\cos (2\phi - 2\Psi_{2} ) \right] , \label{eq:single}
\end{equation}
where $v_{2}(p)$ is the elliptic flow at a given transverse momentum $p=|\vec{p}|$. $\Psi_{2}$ is the event plane, which we further put to zero.

\subsection{Two particles}

Following calculations presented, e.g., in Refs.  \cite{Borghini:2000cm,Borghini:2003ur,Borghini:2006yk,Chajecki:2008vg,Chajecki:2008yi,Bzdak:2010fd,Bzdak:2017zok}, 
the two-particle distribution with TMC is given by
\begin{equation}
f_{2}(p_{1},\phi _{1},p_{2},\phi _{2})=f(p_{1},\phi _{1})f(p_{2},\phi _{2})%
\frac{N}{N-2}\exp \left( -\frac{(p_{1,x}+p_{2,x})^{2}}{2(N-2)\left\langle
p_{x}^{2}\right\rangle _{F}}-\frac{(p_{1,y}+p_{2,y})^{2}}{2(N-2)\left\langle
p_{y}^{2}\right\rangle _{F}}\right) , \label{eq:f2}
\end{equation}
where $p_{x}=p\cos(\phi )$, $p_{y}=p\sin(\phi)$ and using Eq. (\ref{eq:single}) we have
\begin{eqnarray}
\left\langle p_{x}^{2}\right\rangle _{F} &=&\frac{1}{2}\left\langle
p^{2}\right\rangle _{F}\left( 1+\bar{\bar{v}}_{2,F}\right) ,  \notag \\
\left\langle p_{y}^{2}\right\rangle _{F} &=&\frac{1}{2}\left\langle
p^{2}\right\rangle _{F}\left( 1-\bar{\bar{v}}_{2,F}\right) ,
\end{eqnarray}
where
\begin{equation}
\bar{\bar{v}}_{2,F}=\frac{\left\langle v_{2}(p)p^{2}\right\rangle _{F}}{%
\left\langle p^{2}\right\rangle _{F}}=\frac{\int_{F}g(p)v_{2}(p)p^{2}d^{2}p}{%
\int_{F}g(p)p^{2}d^{2}p}.
\end{equation}
The integrations over the full phase space are always denoted by $F$.

Our goal is to calculate
\begin{equation}
\langle e^{2i(\phi _{1}-\phi _{2})}\rangle |_{p_{1},p_{2}}=\frac{%
\int_{0}^{2\pi }f_{2}(p_{1},\phi _{1};p_{2},\phi _{2})e^{2i(\phi _{1}-\phi
_{2})}d\phi _{1}d\phi _{2}}{\int_{0}^{2\pi }f_{2}(p_{1},\phi _{1};p_{2},\phi
_{2})d\phi _{1}d\phi _{2}}=\frac{U_2}{D_2},
\end{equation}%
where $\langle e^{2i(\phi _{1}-\phi _{2})}\rangle $ is calculated
at a given transverse momenta $p_{1}$ and $p_{2}$.

To calculate the numerator we expand $\exp (-A)\approx 1-A+A^{2}/2$ and
neglect all higher terms in Eq. (\ref{eq:f2}). As shown in Ref. \cite{Bzdak:2017zok} the first contribution from
TMC, which is not vanishing at $v_{2}=0$, appears in $A^{2}/2.$ We obtain%
\footnote{%
We skip $\frac{g(p_{1})}{2\pi }\frac{g(p_{2})}{2\pi }\frac{N}{N-2}$
appearing in Eq. (\ref{eq:f2}) since it cancels in the ratio $U_{2}/D_{2}$.} 
\begin{eqnarray}
\frac{U_2}{4\pi ^{2}} &=&v_{2}(p_{1})v_{2}(p_{2})-\frac{%
p_{1}^{2}v_{2}(p_{2})[2v_{2}(p_{1})-\bar{\bar{v}}%
_{2,F}]+p_{2}^{2}v_{2}(p_{1})[2v_{2}(p_{2})-\bar{\bar{v}}_{2,F}]}{%
2(N-2)\left\langle p^{2}\right\rangle _{F}[1-(\bar{\bar{v}}_{2,F})^{2}]}+ 
\notag \\
&&\frac{p_{1}^{4}v_{2}(p_{2})[v_{2}(p_{1})\{4+3(\bar{\bar{v}}_{2,F})^{2}\}-4%
\bar{\bar{v}}_{2,F}]+p_{2}^{4}v_{2}(p_{1})[v_{2}(p_{2})\{4+3(\bar{\bar{v}}%
_{2,F})^{2}\}-4\bar{\bar{v}}_{2,F}]}{8(N-2)^{2}\left\langle
p^{2}\right\rangle _{F}^{2}[1-(\bar{\bar{v}}_{2,F})^{2}]^{2}}+  \notag \\
&&\frac{2p_{1}^{2}p_{2}^{2}[4v_{2}(p_{1})v_{2}(p_{2})\{2+(\bar{\bar{v}}%
_{2,F})^{2}\}-6\bar{\bar{v}}_{2,F}\{v_{2}(p_{1})+v_{2}(p_{2})\}+(\bar{\bar{v}%
}_{2,F})^{2}]}{8(N-2)^{2}\left\langle p^{2}\right\rangle _{F}^{2}[1-(\bar{%
\bar{v}}_{2,F})^{2}]^{2}}+  \notag \\
&&\frac{p_{1}^{2}p_{2}^{2}}{2(N-2)^{2}\left\langle p^{2}\right\rangle
_{F}^{2}[1-(\bar{\bar{v}}_{2,F})^{2}]^{2}}. \label{eq:U2full}
\end{eqnarray}

To calculate the denominator it is enough to take the first term, $\exp
(-A)\approx 1$, since the next terms are suppressed by the powers of $1/N$.
In this case we obtain 
\begin{equation}
D_{2}=4\pi ^{2},
\end{equation}%
and the first correction (assuming $v_{2}^2 \ll 1$) is given by $-4\pi ^{2}\frac{p_{1}^{2}+p_{2}^{2}}{(N-2)\langle
p^{2}\rangle _{F}}$.

The last term of $U_2$ in Eq. (\ref{eq:U2full}), discussed in Ref. \cite{Bzdak:2017zok}, is driven by momentum
conservation and it does not vanish for $v_{2}=0$. It scales like $1/N^{2}$.
The third and the fourth terms of $U$ are suppressed also by $1/N^{2}$ and
additionally they are multiplied by $v_{2}^{2}$, and thus can be neglected 
(unless one of $p_{i}$ is very small). 
Moreover we can use $1-(\bar{\bar{v}}_{2,F})^{2}\approx 1$ etc. With a good approximation we obtain
\begin{equation}
c_{2}\{2\}\approx v_{2}(p_{1})v_{2}(p_{2})-\frac{%
p_{1}^{2}v_{2}(p_{2})[2v_{2}(p_{1})-\bar{\bar{v}}%
_{2,F}]+p_{2}^{2}v_{2}(p_{1})[2v_{2}(p_{2})-\bar{\bar{v}}_{2,F}]}{%
2(N-2)\left\langle p^{2}\right\rangle _{F}}+\frac{p_{1}^{2}p_{2}^{2}}{%
2(N-2)^{2}\left\langle p^{2}\right\rangle _{F}^{2}}, \label{eq:U2appr}
\end{equation}%
where by definition $c_{2}\{2\} = \frac{U_2}{D_2} = \langle e^{2i(\phi _{1}-\phi _{2})}\rangle |_{p_{1},p_{2}}$.

Finally, for $p_{1}=p_{2}=p$ we have
\begin{equation}
c_{2}\{2\}\approx (v_{2}(p))^{2}-\frac{p^{2}v_{2}(p)[2v_{2}(p)-\bar{%
\bar{v}}_{2,F}]}{(N-2)\left\langle p^{2}\right\rangle _{F}}+\frac{p^{4}}{%
2(N-2)^{2}\left\langle p^{2}\right\rangle _{F}^{2}}. \label{eq:U2appr2}
\end{equation}%

As seen from Eq. (\ref{eq:U2appr2}), for large $N$ the value of $c_{2}\{2\}$ is 
driven by $(v_{2}(p))^{2}$ and for small $N$ the last term (TMC contribution) becomes increasingly important.

\subsection{Four particles}

The four particle density with TMC is given by
\begin{eqnarray}
f_{4}(p_{1},\phi _{1},...,p_{4},\phi _{4}) &=&f(p_{1},\phi _{1})\cdots
f(p_{4},\phi _{4})\frac{N}{N-4}\times  \notag \\
&&\exp \left( -\frac{(p_{1,x}+...+p_{4,x})^{2}}{2(N-4)\left\langle
p_{x}^{2}\right\rangle _{F}}-\frac{(p_{1,y}+...+p_{4,y})^{2}}{%
2(N-4)\left\langle p_{y}^{2}\right\rangle _{F}}\right) ,\label{eq:f4}
\end{eqnarray}%
and
\begin{equation}
\langle e^{2i(\phi _{1}+\phi _{2}-\phi _{3}-\phi _{4})}\rangle
|_{p_{1},p_{2},p_{3},p_{4}}=\frac{\int_{0}^{2\pi }f_{4}(p_{1},\phi
_{1},...,p_{4},\phi _{4})e^{2i(\phi _{1}+\phi _{2}-\phi _{3}-\phi
_{4})}d\phi _{1}\cdots d\phi _{4}}{\int_{0}^{2\pi }f_{4}(p_{1},\phi
_{1},...,p_{4},\phi _{4})d\phi _{1}\cdots d\phi _{4}}=\frac{U_4}{D_4} .
\end{equation}

In this case we expand $\exp (-A)$ up to the fourth order, $\exp (-A)\approx
1-A+\frac{A^{2}}{2}-\frac{A^{3}}{6}+\frac{A^{4}}{24}$ in Eq. (\ref{eq:f4}) since, as shown in
\cite{Bzdak:2017zok}, the first non-vanishing TMC term at $v_{2}=0$ is coming from 
$A^{4}/24$. The results for arbitrary $p_{1},p_{2},p_{3}$ and $p_{4}$ are too
complicated to include in the paper and in the following we assume $p_{i}=p$, $i=1,2,3,4$. 
Even in this case, the resulting $U_4$ is rather complicated and we present it in the Appendix.
As in the previous Section, to calculate the denominator, $D_4$, it is enough to take the first term, $\exp
(-A)\approx 1$, resulting in $D_4 = 16\pi ^{4}$, since the next terms are suppressed by the powers of $1/N$ (the first correction, assuming $v_{2}^2 \ll 1$, to $D_4$ is given by $-16\pi ^{4}\frac{4p^{2}}{(N-4)\langle p^{2}\rangle _{F}}$).

Using $(v_2(p))^{2}\ll 1$, $(\bar{\bar{v}}_{2,F})^{2} \ll 1$ etc., we obtain (see the Appendix for details)
\begin{eqnarray}
c_{2}\{4\} &\approx &(v_{2}(p))^{4}-\frac{%
2p^{2}(v_{2}(p))^{3}[2v_{2}(p)-\bar{\bar{v}}_{2,F}]}{(N-4)\left\langle
p^{2}\right\rangle _{F}}+\frac{2p^{4}(v_{2}(p))^{2}}{(N-4)^{2}\left\langle
p^{2}\right\rangle _{F}^{2}}-\frac{2p^{6}v_{2}(p)[8v_{2}(p)-3\bar{\bar{v}}%
_{2,F}]}{(N-4)^{3}\left\langle p^{2}\right\rangle _{F}^{3}}  \notag \\
&&+\frac{p^{8}[442(v_{2}(p))^{2}-360v_{2}(p)\bar{\bar{v}}_{2,F}+27(\bar{\bar{%
v}}_{2,F})^{2}]}{6(N-4)^{4}\left\langle p^{2}\right\rangle _{F}^{4}}+\frac{%
3p^{8}}{2(N-4)^{4}\left\langle p^{2}\right\rangle _{F}^{4}} - 2(c_{2}\{2\})^{2}, \label{eq:U4appr}
\end{eqnarray}%
where $c_{2}\{2\}$ is given in Eq. (\ref{eq:U2appr2}).

\section{Results} 

Here we present the results for $c_2\{2\}$ and $c_2\{4\}$ based respectively on Eqs. (\ref{eq:U2appr2}) and (\ref{eq:U4appr}). We checked that practically identical results are obtained with the full formulas, Eqs. (\ref{eq:U2full}) and (\ref{eq:U4full}). 

\begin{figure}[t]
\begin{center}
\includegraphics[scale=0.4]{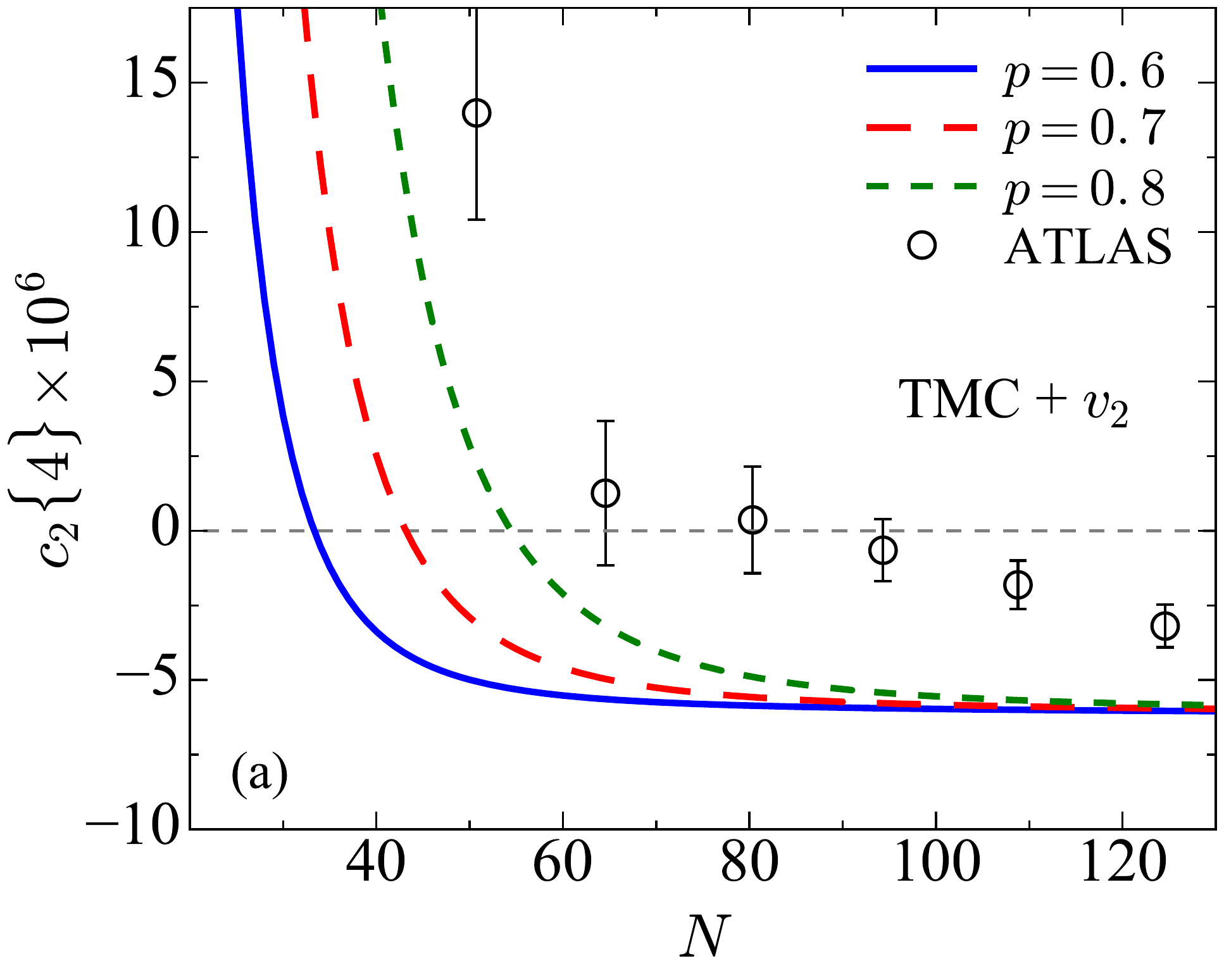}\hspace{6mm} %
\includegraphics[scale=0.4]{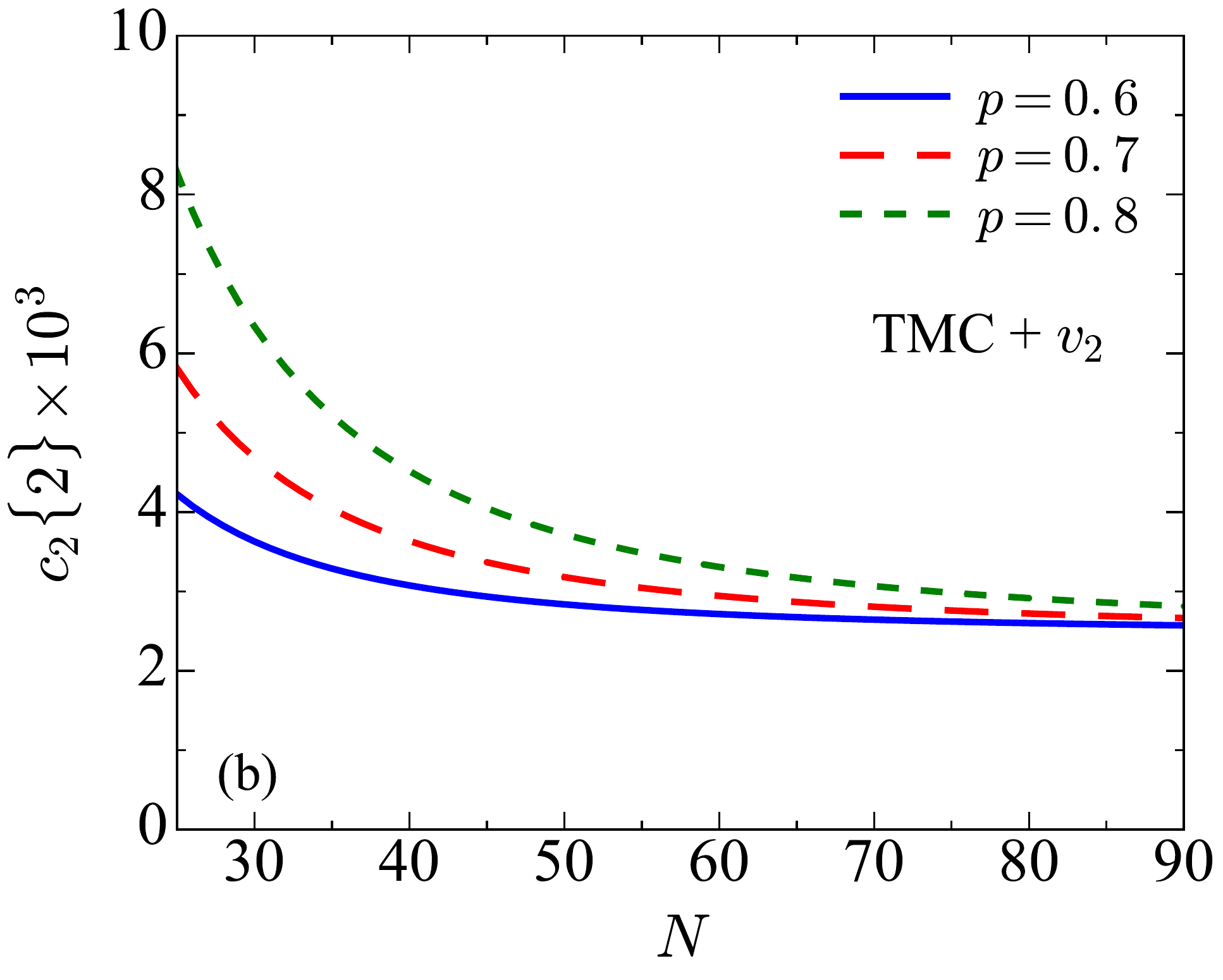}
\end{center}
\par
\vspace{-5mm}
\caption{(a) $c_{2}\{4\}$ and (b) $c_{2}\{2\}$ from transverse momentum conservation 
in the presence of hydro-like elliptic flow as a function of the number
of produced particles $N$ for various values of transverse momenta $p$. In
this calculation $v_2=0.05$ and $\bar{\bar{v}}_{2,F}=0.025$ (our results very weakly
depends on $\bar{\bar{v}}_{2,F}$) and $\langle p^{2}\rangle_{F}=0.5^{2}$. 
$c_{2}\{k\}$ depends only on $p^{2}/\langle p^{2}\rangle _{F}$ and thus the presented 
three curves correspond to $p^{2}/\langle p^{2}\rangle _{F}=1.44,$ $1.96,$ $2.56$.
For an order of magnitude comparison (see text for details) we also show 
the ATLAS results \cite{Aaboud:2017blb} 
for $c_{2}\{4\}$ from the 13 TeV proton-proton data
using the three-subevent cumulant method.}
\label{fig:1}
\end{figure}

\begin{figure}[t]
\begin{center}
\includegraphics[scale=0.4]{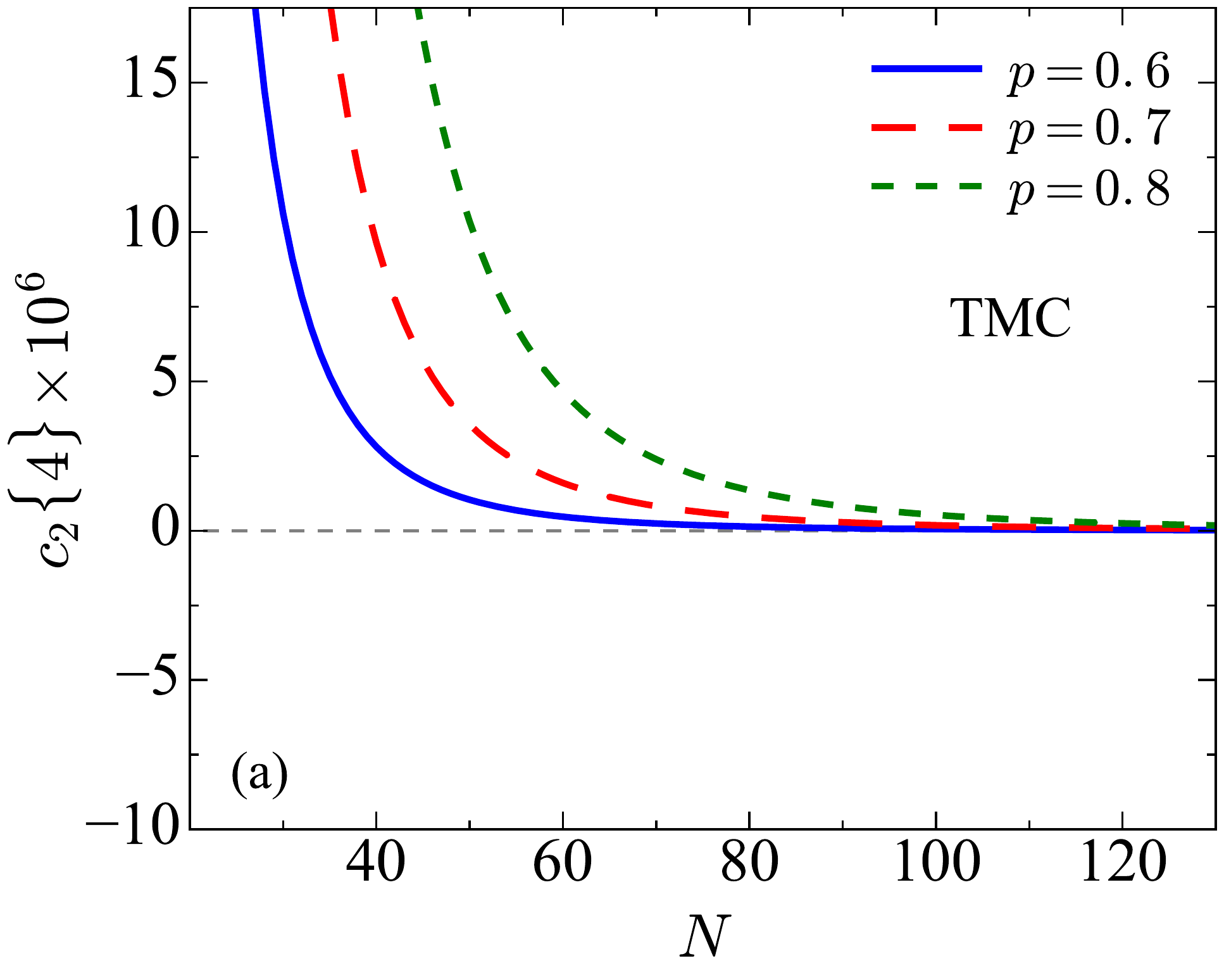}\hspace{6mm} %
\includegraphics[scale=0.4]{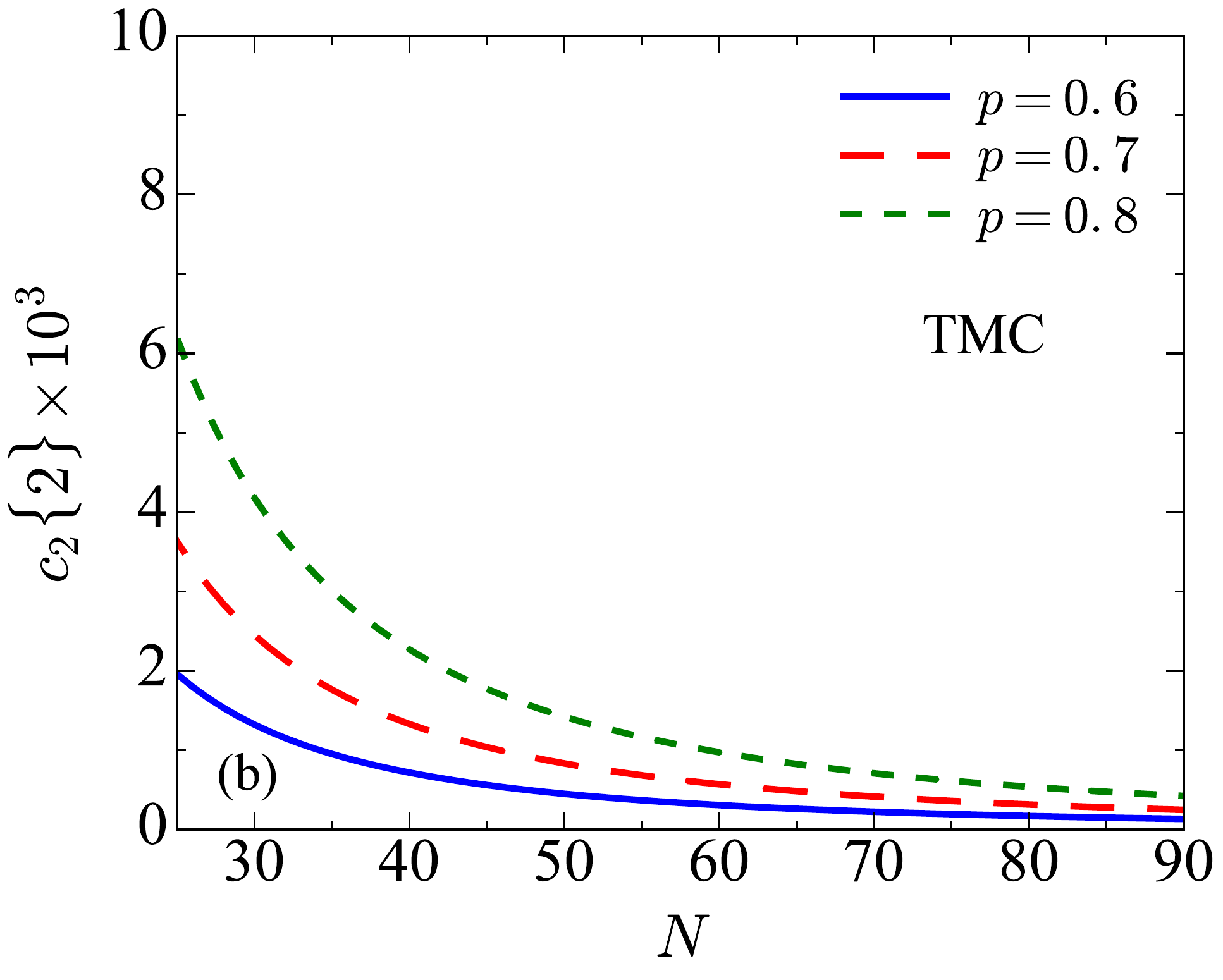}
\end{center}
\par
\vspace{-5mm}
\caption{(a) $c_{2}\{4\}$ and (b) $c_{2}\{2\}$ from transverse momentum conservation 
only as a function of the number of produced particles $N$ for various values of 
transverse momenta $p$. In this calculation $\langle p^{2}\rangle_{F}=0.5^{2}$.}
\label{fig:2}
\end{figure}

In Fig. \ref{fig:1} we present $c_{2}\{4\}$ in panel (a) and $c_{2}\{2\}$ in panel (b) as a function of the number
of produced particles $N$ for various values of transverse momenta $p=0.6$, $0.7$ and $0.8$. 
In this calculation $v_2=0.05$ and $\bar{\bar{v}}_{2,F}=0.025$ (our results very weakly
depends on $\bar{\bar{v}}_{2,F}$), and $\langle p^{2}\rangle_{F}=0.5^{2}$. 
$c_{2}\{k\}$ depends only on $p^{2}/\langle p^{2}\rangle _{F}$ and thus the presented three curves correspond to 
$p^{2}/\langle p^{2}\rangle _{F}=1.44,$ $1.96,$ $2.56$. In Fig. \ref{fig:2} we present results with $v_2=\bar{\bar{v}}_{2,F}=0$, namely there is no contribution from the hydro-like elliptic flow and only TMC contributes to the signal. This case was discussed in detail in our previous paper \cite{Bzdak:2017zok}.

Our main observation is that  
$c_{2}\{4\}$ changes sign as a function of $N$ in the presence of hydro-like elliptic flow.
This can be easily understood: For large $N$, $c_{2}\{4\}$ is dominated by $-(v_2(p))^4$ and for small $N$ the positive contribution from TMC \cite{Bzdak:2017zok} 
becomes dominant.  
The calculated cumulants are functions of many variables, i.e., $N$, $v_2(p)$, $\bar{\bar{v}}_{2,F}$, $p$ 
and $\langle p^{2}\rangle_{F}$. However, $c_{2}\{2\}$ and $c_{2}\{4\}$ depend only on 
$p^{2}/\langle p^{2}\rangle _{F}$, 
and very weakly depend on $\bar{\bar{v}}_{2,F}$. There is a rather strong dependence on $v_2(p)$ resulting from the leading contributions $(v_2(p))^2$ and $(v_2(p))^4$ to $c_{2}\{2\}$ and $c_{2}\{4\}$, respectively. 

Note that in our results, presented in Fig. (\ref{fig:1}), we assumed a constant $v_2(p)$ as a function of $N$. 
However, assuming that $v_2(p)$ decreases with 
decreasing $N$ does not effect our conclusions but results in 
different $N$ for which $c_2\{4\}$ changes its sign. 

Finally, we note that in our calculations we assumed that $c_2\{2\}$ and $c_2\{4\}$ are effected only by TMC and the hydro-like elliptic flow, $v_2(p)$, see Eq. (\ref{eq:single}). Consequently, the recent ATLAS measurement of multi-particle azimuthal correlations with the subevent cumulant method, which is expected to reduce significantly short-range non-flow components \cite{Jia:2017hbm}, is presumably more applicable to examine our results. That is why in Fig. \ref{fig:1} we qualitatively compare our results with the ATLAS data \cite{Aaboud:2017blb} for $c_{2}\{4\}$ from the 13 TeV proton-proton data using the three-subevent cumulant method.\footnote{However, this should be considered at best as an order of magnitude comparison. We assumed that our $N$ corresponds to $\frac{3}{2}\langle N_{\rm ch} \rangle$, where  $\langle N_{\rm ch} \rangle$ is the number of charged particles in $|\eta|<2.5$ and $0.3<p_{}<3$ GeV as defined by ATLAS \cite{Aaboud:2017blb}, which is not necessarily the case. This is not unjustified since the transverse momentum is expected to be conserved locally (see, e.g., Ref. \cite{Pratt:2010zn}) however, this problem requires more sophisticated studies and goes beyond the scope of the present paper. Moreover, we calculated at fixed $p_i=p$ ($i=1,2,3,4$) and in the ATLAS data, shown in Fig. \ref{fig:1}, transverse momenta are integrated over $0.3<p<3$ GeV.}

\section{Conclusions}

In this paper we calculated analytically the multi-particle azimuthal correlations, $c_2\{2\}$ and $c_2\{4\}$, originating from the global conservation of transverse momentum in the presence of hydro-like elliptic flow. Comparing to our previous calculation of transverse momentum conservation only, the presence of the elliptic flow leads to an enhancement of $c_2\{2\}$ and suppression of $c_2\{4\}$, which can naturally explain the sign change of $c_2\{4\}$ at a small number of produced particles. This is qualitatively consistent with the recent ATLAS measurement of multi-particle azimuthal correlations with the subevent cumulant method. Our results offer a new insight into the problem of the onset of collectivity in small systems.


\vspace{\baselineskip} 
\noindent\textbf{Acknowledgments} 
\newline
{} 
A.B. is partially supported by the Faculty of Physics and Applied Computer Science 
AGH UST statutory tasks No. 11.11.220.01/1 within subsidy of Ministry of Science 
and Higher Education, and by the National Science Centre, Grant No. DEC-2014/15/B/ST2/00175.
G.-L.M. is supported by the Major State Basic Research Development Program in China under Grant No. 2014CB845404, the National Natural Science Foundation of China under Grants No. 11522547, 11375251, and 11421505.

\appendix
\section{}

Expanding $\exp (-A)$ up to the fourth order in Eq. (\ref{eq:f4}) we obtain
\begin{equation}
\frac{U_4}{16\pi ^{4}}=U^{(0)}_{4}+U^{(1)}_{4}+U^{(2)}_{4}+U^{(3)}_{4}+U^{(4)}_{4}, \label{eq:U4full}
\end{equation}
where
\begin{eqnarray}
U^{(0)}_{4} &=&(v_{2}(p))^{4}, \\
U^{(1)}_{4} &=&-\frac{2p^{2}(v_{2}(p))^{3}[2v_{2}(p)-\bar{\bar{v}}_{2,F}]}{%
(N-4)\left\langle p^{2}\right\rangle _{F}[1-(\bar{\bar{v}}_{2,F})^{2}]}, \\
U^{(2)}_{4} &=&\frac{p^{4}(v_{2}(p))^{2}[4+28(v_{2}(p))^{2}-40v_{2}(p)\bar{\bar{v%
}}_{2,F}+5(\bar{\bar{v}}_{2,F})^{2}+15(v_{2}(p)\bar{\bar{v}}_{2,F})^{2}]}{%
2(N-4)^{2}\left\langle p^{2}\right\rangle _{F}^{2}[1-(\bar{\bar{v}}%
_{2,F})^{2}]^{2}},
\end{eqnarray}%
and
\begin{eqnarray}
U^{(3)}_{4} &=&-\frac{p^{6}v_{2}(p)}{6(N-4)^{3}\left\langle p^{2}\right\rangle
_{F}^{3}[1-(\bar{\bar{v}}_{2,F})^{2}]^{3}}\left[ 24v_{2}(p)\{4+11(\bar{\bar{v%
}}_{2,F})^{2}\}-9\bar{\bar{v}}_{2,F}\{4+(\bar{\bar{v}}_{2,F})^{2}\}\right. 
\notag \\
&&\left. +16(v_{2}(p))^{3}\{16+27(\bar{\bar{v}}_{2,F})^{2}\}-3(v_{2}(p))^{2}%
\bar{\bar{v}}_{2,F}\{234+61(\bar{\bar{v}}_{2,F})^{2}\}\right] ,
\end{eqnarray}%
\begin{eqnarray}
U^{(4)}_{4} &=&\frac{p^{8}}{96(N-4)^{4}\left\langle p^{2}\right\rangle
_{F}^{4}[1-(\bar{\bar{v}}_{2,F})^{2}]^{4}}\left[ 8(v_{2}(p))^{2}\{884+4377(%
\bar{\bar{v}}_{2,F})^{2}+619(\bar{\bar{v}}_{2,F})^{4}\}\right.  \notag \\
&&\left. -1440v_{2}(p)\bar{\bar{v}}_{2,F}\{4+3(\bar{\bar{v}}_{2,F})^{2}\}+54(%
\bar{\bar{v}}_{2,F})^{2}\{8+(\bar{\bar{v}}_{2,F})^{2}\}\right.  \notag \\
&&\left. -32(v_{2}(p))^{3}\bar{\bar{v}}_{2,F}\{1502+1193(\bar{\bar{v}}%
_{2,F})^{2}\}+(v_{2}(p))^{4}\{10880+38928(\bar{\bar{v}}_{2,F})^{2}+5163(\bar{%
\bar{v}}_{2,F})^{4}\}\right] +  \notag \\
&&\frac{3p^{8}}{2(N-4)^{4}\left\langle p^{2}\right\rangle _{F}^{4}[1-(\bar{%
\bar{v}}_{2,F})^{2}]^{4}}.
\end{eqnarray}

Assuming $(v_2(p))^{2}\ll 1$, $(\bar{\bar{v}}_{2,F})^{2} \ll 1$ etc., taking $D_4 = 16\pi ^{4}$, and using Eq. (\ref{eq:c24-def}) we obtain $c_2\{4\}$ given in Eq. (\ref{eq:U4appr}).

\end{document}